# Low-level RF

**Part I: Longitudinal dynamics and beam-based loops in synchrotrons**

*P. Baudrenghien*
CERN, Geneva, Switzerland

**Abstract**
The low-level RF system (LLRF) generates the drive sent to the high-power equipment. In synchrotrons, it uses signals from beam pick-ups (radial and longitudinal) to minimize the beam losses and provide a beam with reproducible parameters (intensity, bunch length, average momentum and momentum spread) for either the next accelerator or the physicists. This presentation is the first of three: it considers synchrotrons in the low-intensity regime where the voltage in the RF cavity is not influenced by the beam. As the author is in charge of the LHC LLRF and currently commissioning it, much material is particularly relevant to hadron machines. A section is concerned with radiation damping in lepton machines.

## 1   Applied longitudinal dynamics in synchrotrons

Synchrotrons are circular accelerators whose RF frequency varies during the acceleration ramp to keep the particles on a centred orbit. In this section we study the dynamics of a particle that periodically crosses the accelerating cavities and gains or loses energy by interaction with the electric field. The intent is to cover the basics of longitudinal dynamics, required to understand low level RF (LLRF). Please consult Refs. [1]–[8] for a more detailed coverage.

### 1.1   The synchronous particle

We first consider a reference particle that stays exactly on the centred orbit turn after turn. This fictitious particle is called the **synchronous particle**.

The RF frequency $f_{RF}$ must be locked to the revolution frequency $f_{rev}$ of the synchronous particle to have a coherent effect turn after turn. The ratio (integer $h$) is called the harmonic number

$$f_{RF} = h \cdot f_{rev} \tag{1}$$

$$f_{RF} = h \cdot \frac{v}{2\pi R_0} = \frac{hc}{2\pi R_0} \beta \tag{2}$$

$$\beta = \frac{v}{c} \tag{3}$$

with $2\pi R_0$ the machine circumference and $v$ the speed of the particle.

In order for the synchronous particle to stay exactly on the centred orbit, the radial component of the magnetic force must compensate the centrifugal force. Let $\rho$ be the bending radius of the magnet, and $q$ the charge of the particle, we then have

$$\frac{m \cdot v^2}{\rho} = q \cdot v \cdot B \quad , \tag{4}$$

$$p = q.\rho.B \quad . \tag{5}$$

Using the relations between $\beta$ (ratio of particle velocity to the velocity of light), $p$ (momentum), and $\gamma$ (ratio of particle total energy $E$ to the rest energy $E_0$) we get (see Appendix A)

$$f_{RF} = \frac{hc}{2\pi R_0}\beta = \frac{hc}{2\pi R_0}\frac{1}{\sqrt{1+\left(\frac{E_0}{c.p}\right)^2}} = f_\infty\sqrt{1-\frac{1}{\gamma^2}} \quad , \tag{6}$$

with the RF frequency at infinite energy

$$f_\infty = \frac{hc}{2\pi R_0} \quad . \tag{7}$$

Using the linear relation between the momentum and the dipole field — Eq. (5) — Eq. (6) can be rewritten

$$f_{RF} = f_\infty \frac{B}{\sqrt{B^2 + \left(\frac{1}{c.\rho}\frac{E_0}{q}\right)^2}} \quad . \tag{8}$$

Let us now analyse Eqs. (6) and (8):

- The $f_{RF}$ vs $B$ relation is non-linear.
- The frequency swing depends on the range of $\gamma$ from injection to extraction. We have a large frequency swing when the injection energy is low so that the speed varies greatly during the ramp (non-relativistic machine).
- For highly relativistic machines (electrons) the RF frequency can be kept constant.
- Low-energy proton or ion machines will have a large frequency swing.
- Heavy ions have a larger $E_0/q$ ratio than protons because neutrons have no charge. If accelerated with the same magnetic ramp, the frequency swing will be larger.
- If the frequency swing is large, the RF frequency would best be controlled from a measurement of the dipole field.
- It is the responsibility of the LLRF to make the RF frequency track the dipole field according to Eq. (8).

Some examples:

- $e^+e^-$ ($E_0 = 0.511$ MeV) acceleration in the SPS as LEP injector, from 3 GeV/$c$ to 22 GeV/$c$ at constant frequency 200.395 MHz.
- Proton ($E_0 = 938.26$ MeV) acceleration in the LHC from 450 GeV/$c$ (400.788860 MHz) to 3.5 TeV/$c$ (400.789713 MHz).
- Original proton acceleration in the CPS (1959, $h = 20$) from 50 MeV/$c$ (2.9 MHz) to 25 GeV/$c$ (9.54 MHz).
- Lead ion $^{208}$Pb$^{82+}$ acceleration in the SPS from 5.87 GeV/u (kinetic energy per nucleon) at 198.501 MHz to 160 GeV/u (200.393 MHz) for injection in the LHC.

Figure 1 shows the LHC frequency ramp used at the beginning of 2010 for protons. By the end of the year the ramp was shortened to 15 minutes. The frequency swing is less than 1 kHz at 400 MHz.

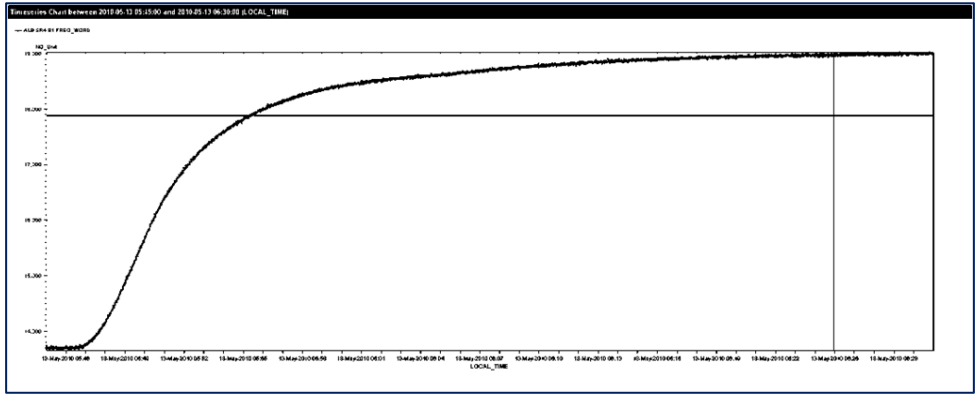

**Fig. 1:** The 45 minute long LHC frequency ramp from 450 GeV/*c* (400.788 860 MHz) to 3.5 TeV/*c* (400.789 713 MHz) used at the beginning of 2010

Let us now consider the phase $\phi_s$ of the RF when the synchronous particle crosses the electric field. This phase is called **synchronous** or **stable phase**. The energy increase per turn, caused by the electric field is

$$\Delta E_{turn} = q\, V \sin \varphi_s \quad . \tag{9}$$

The interaction with the electric field takes place at each turn. Assuming that the timescale of longitudinal dynamics is much longer than a revolution period, discrete interactions can be approximated by continuous-time derivatives and we get

$$\frac{1}{f_{rev}} \frac{dE}{dt} = q\, V \sin \varphi_s \quad . \tag{10}$$

Using the linear relation between energy and momentum (Appendix A)

$$2\pi R_0 \frac{dp}{dt} = q\, V \sin \varphi_s \quad . \tag{11}$$

The LHS is defined by the machine momentum ramp. That, in turn, defines the product $V \sin \phi_s$

– in hadron colliders $dp/dt = 0$ and the stable phase is zero or 180 degrees,
– in ramping synchrotrons, $\phi_s$ is chosen to give the desired **bucket area** (Section 1.3).

## 1.2 Useful differential relations

The previous section showed that the synchronous RF frequency and consequently the revolution frequency of the synchronous particle must track the *B* field to keep the beam centred, Eq. (8). This corresponds to imposing the average radius of the particle trajectory *R (R = R₀)* and the dipole field *B*, and deriving $f_{rev}$ (or $f_{RF}$). Of the four variables *(f, B, p, R)*, only two are independent for the synchronous particle. The relationship is non-linear but it can be linearized locally. This leads to four very useful differential relations [3]

$$p = p(R, B) \Rightarrow \frac{\Delta p}{p} = \gamma_t^2 \frac{\Delta R}{R} + \frac{\Delta B}{B} \quad , \tag{12}$$

$$p = p(f, R) \Rightarrow \frac{\Delta p}{p} = \gamma^2 \frac{\Delta f}{f} + \gamma^2 \frac{\Delta R}{R} \quad , \tag{13}$$

$$B = B(f, p) \Rightarrow \frac{\Delta B}{B} = \gamma_t^2 \frac{\Delta f}{f} + \frac{\gamma^2 - \gamma_t^2}{\gamma^2} \frac{\Delta p}{p} \quad , \tag{14}$$

$$B = B(f,R) \Rightarrow \frac{\Delta B}{B} = \gamma^2 \frac{\Delta f}{f} + \left(\gamma^2 - \gamma_t^2\right)\frac{\Delta R}{R} \quad . \tag{15}$$

The transition energy $\gamma_t$ will be presented shortly. Let us now use the above relations.

- **Matching the magnetic field at injection:** We measure the radial displacement on first turn $\Delta R$ and wish to trim the magnetic field $B$ to centre the beam. Since the momentum is fixed (defined by the injector), we will use Eq. (12), with $\Delta p = 0$, to derive the appropriate $\Delta B$ from the measured $\Delta R$:

$$\frac{\Delta B}{B} = -\gamma_t^2 \frac{\Delta R}{R} \quad \text{at } p \text{ constant} \quad . \tag{16}$$

- **Displacing the circulating beam by trimming the RF frequency:** This operation is used routinely for chromaticity measurement. We keep the magnetic field $B$ constant and wish to relate radial displacement $\Delta R$ with the frequency trim $\Delta f$. We can use Eq. (15), setting $\Delta B = 0$. This gives the desired Hz/mm scaling factor

$$\frac{\Delta f}{f} = \left(\frac{\gamma_t^2}{\gamma^2} - 1\right)\frac{\Delta R}{R} \quad \text{at B constant} \quad . \tag{17}$$

Or we can relate the frequency trim to a momentum offset using Eq. (14) with $B$ constant

$$\frac{\Delta f}{f} = \left(\frac{1}{\gamma^2} - \frac{1}{\gamma_t^2}\right)\frac{\Delta p}{p} = \eta \frac{\Delta p}{p} \quad \text{at } B \text{ constant} \quad . \tag{18}$$

Here $\eta$ is called the **slippage factor**. It changes sign at the transition energy. At constant magnetic field, if the momentum is increased, both particle mass and speed will increase. An increase of mass drives the particle on an outer orbit, therefore reducing the revolution frequency as the trajectory is longer. On the other hand, an increase of particle speed always tends to increase the revolution frequency as the particle travels faster.

At low energy the effect of the particle speed dominates and the revolution frequency increases with momentum (positive $\eta$). At high energy the speed barely changes and the lengthening of the orbit dominates. The revolution frequency decreases with momentum (negative $\eta$). At **transition energy** the two effects compensate and the revolution frequency becomes insensitive to momentum.

When using a formula including the slippage factor, beware that some authors use revolution period instead of revolution frequency in the definition (18). The resulting η has the same absolute value but **inverted sign**. So check the definition.

### 1.3   Non-synchronous particles

So far we have considered the synchronous particle: it has the correct momentum — Eq. (5) — so that it stays exactly on the centred orbit turn after turn. The RF frequency is an integer multiple of the synchronous particle revolution frequency so that this fictitious particle crosses the electric field at a constant phase $\phi_s$, turn after turn. In this section we now consider a particle P having a small momentum offset with respect to the synchronous particle. As a consequence it has a different revolution frequency and crosses the cavity at a slightly different RF phase. Let $(p_s, \phi_s)$ refer to the synchronous particle and $(p, \phi)$ refer to particle P. Given the small momentum difference P has also a different revolution frequency

$$\tilde{\varphi} = \varphi - \varphi_s \quad , \tag{19}$$

$$\frac{d\tilde{\varphi}}{dt} = -2\pi h \, \tilde{f}_{rev} \quad . \tag{20}$$

We have a minus sign because $\phi$ is the RF phase when P crosses the cavity. (In this paper the superscript ~ represents deviations with respect to the synchronous particle while the subscript *s* refers to the synchronous particle.) The above relation is kinematic only. Let us now introduce the electric force.

Crossing the cavity at a different RF phase, the momentum increase is different for P and for the synchronous particle

$$2\pi R_0 \frac{dp_s}{dt} = qV \sin\varphi_s \quad , \tag{21}$$

$$2\pi R_0 \frac{dp}{dt} = qV \sin\varphi \quad , \tag{22}$$

$$2\pi R_0 \frac{d\tilde{p}}{dt} = qV \sin\varphi - qV \sin\varphi_s \quad . \tag{23}$$

The slippage factor — Eq. (18) — relates a momentum offset to a frequency offset, at constant magnetic field

$$\frac{d\tilde{\varphi}}{dt} = -2\pi h \, \tilde{f}_{rev} = -2\pi \eta \, h f_{rev} \frac{\tilde{p}}{p_s} \quad . \tag{24}$$

Differentiating Eq. (24) we get

$$\frac{d^2 \tilde{\varphi}}{dt^2} = -\frac{2\pi \eta \, h f_{rev}}{p_s} \frac{d\tilde{p}}{dt} \quad . \tag{25}$$

Now merging the above two equations we get a second-order differential equation describing the **synchrotron motion**. Notice the non-linearity (sine term)

$$\frac{d^2 \tilde{\varphi}}{dt^2} + \frac{\eta f_{RF}}{R_0 p_s} qV \left( \sin\varphi - \sin\varphi_s \right) = 0 \quad . \tag{26}$$

Let us first consider small phase deviations with respect to the synchronous particle

$$\sin\varphi = \sin(\varphi_s + \tilde{\varphi}) = \sin\varphi_s \cos\tilde{\varphi} + \sin\tilde{\varphi}\cos\varphi_s \approx \sin\varphi_s + \tilde{\varphi}\cos\varphi_s \quad . \tag{27}$$

And Eq. (26) becomes linear

$$\frac{d^2 \tilde{\phi}}{dt^2} + \frac{\eta f_{RF} \, qV \cos\phi_s}{R_0 \, p_s} \tilde{\phi} = 0 \tag{28}$$

or

$$\frac{d^2 \tilde{\phi}}{dt^2} + \Omega_s^2 \, \tilde{\phi} = 0 \tag{29}$$

with

$$\Omega_s = \sqrt{\frac{\eta f_{RF} \, qV \cos\varphi_s}{R_0 \, p_s}} \quad . \tag{30}$$

If $\Omega_s^2$ is positive, the equation of synchrotron motion represents an **undamped harmonic oscillator** with resonant frequency $\Omega_s$, called the **synchrotron frequency**. Given a phase or momentum error as initial conditions, the particle will oscillate endlessly around the stable phase, exchanging longitudinal displacement with momentum offset. The period of the synchrotron frequency is the characteristic time-response of the beam in the longitudinal plane. We will call **adiabatic** the evolutions that are slow with respect to this period.

If $\Omega_s^2$ is negative, the solutions of Eq. (29) will be the combination of a decaying and a growing exponential and the motion is unbounded. We are interested in situations where the distance between particle P and the synchronous particle remains bounded and that requires

$$\eta \cos \varphi_s \geq 0 \quad . \tag{31}$$

In that case the motion will be periodic. Recall that the slippage factor $\eta$ is

$$\eta = \frac{\frac{\Delta f}{f}}{\frac{\Delta p}{p}}\bigg|_{B\,cst} = \left(\frac{1}{\gamma^2} - \frac{1}{\gamma_t^2}\right) \quad . \tag{32}$$

The sign of $\cos \phi_s$ therefore changes at transition. We have

- Acceleration below transition

$$\gamma \leq \gamma_t \Rightarrow \eta \geq 0 \Rightarrow \cos \varphi_s \geq 0 \Rightarrow \varphi_s \in \left[0, \frac{\pi}{2}\right] \quad . \tag{33}$$

- Acceleration above transition

$$\gamma \geq \gamma_t \Rightarrow \eta \leq 0 \Rightarrow \cos \varphi_s \leq 0 \Rightarrow \varphi_s \in \left[\frac{\pi}{2}, \pi\right] \quad . \tag{34}$$

Let us return to the synchrotron motion Eq. (26), before linearization. After a first integration, it becomes

$$\frac{1}{2}\left(\frac{\frac{d\tilde{\varphi}}{dt}}{\Omega_s}\right)^2 - \frac{\left(\cos \tilde{\varphi} + \tilde{\varphi} \sin \varphi_s\right)}{\cos \varphi_s} = C \quad . \tag{35}$$

For each value of the constant *C* we have a different trajectory. Figure 2 shows a phase space representation of these trajectories.

Analysis

- For small deviations from the stable phase the trajectories are circular in phase space. This corresponds to the linearized Eq. (29). For larger deviations the trajectories are deformed, but still closed, corresponding to a **quasi-harmonic undamped oscillator.** Closed trajectories (stable motion) are marked in blue on Fig. 2.

- Above some excursion the trajectories are not closed anymore and these particles are not controlled by the RF (green traces). The limiting closed trajectory is called the **separatrix** marked in red on the figure. The enclosed surface in phase space is called the **bucket area.**

- If there is no acceleration (Fig. 2, top left) the particles outside the separatrix drift in the machine, 'surfing' over the buckets. Such a situation is found during injection, when some particles fall outside the buckets and are not **captured** by the RF. They are called **unbunched beam**.

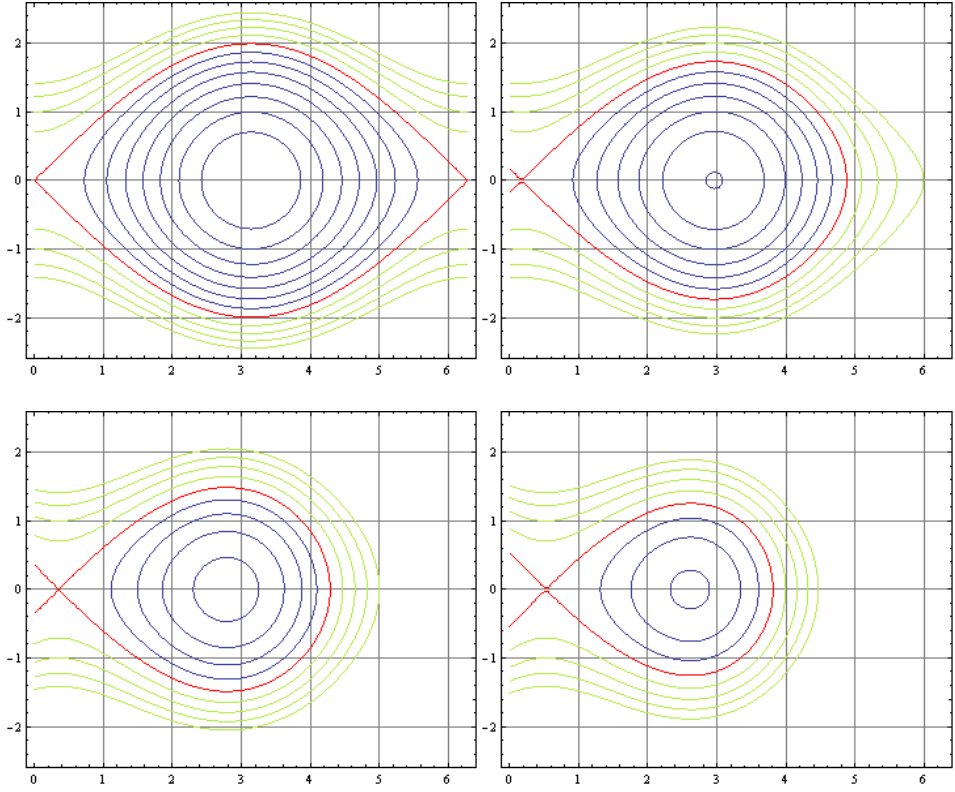

**Fig. 2:** Trajectories in **normalized phase space** ($\phi$, $1/\Omega_s\, d\phi/dt$) above transition for synchronous phase 180 degrees (top left), 170 (top right), 160 (bottom left) and 150 degrees (bottom right). The separatrix is in red. Stable trajectories are shown in blue, unstable motion appears in green. The particles move clockwise on the trajectories.

- The previous phase space plots are in normalized ($\phi$, $1/\Omega_s (d\phi/dt)$) units. The trajectories are similar if the horizontal axis is time and the vertical axis is $\tilde{p}$ or $\tilde{E}$ (momentum or energy deviation with respect to synchronism). However, momentum and energy deviations are related to $d\phi/dt$ via the slippage factor $\eta$ that changes sign at transition, Eq. (24). Using these for the y-axis, the phase space trajectories will thus be travelled in the anti-clockwise direction below transition and in the clockwise direction above transition (Fig. 3).

- In the presence of acceleration, the unbunched beam sees its momentum decrease with respect to the synchronous particle as it does not interact with the electric field coherently turn after turn. Considering the correct direction of travel on the trajectories, we see on Fig. 3 that the momentum deviation decreases in all cases. (In reality it is the momentum of the synchronous particle that increases.) As the magnetic field increases, these particles move inwards in the vacuum chamber and are lost.

- The bucket area $A$ is usually expressed in physical energy × time unit (eVs)

$$A = \left( \frac{16\sqrt{q}}{(2\pi)^{3/2}\sqrt{h}} \right) \left( \frac{\beta\sqrt{E_s}}{f_{RF}\sqrt{|\eta|}} \right) \sqrt{V}\, \alpha(\varphi_s) \quad . \tag{36}$$

The function $\alpha(\phi_s)$ is a non-linear function describing the rapid reduction of bucket area with the stable phase (Fig. 2). It is equal to 1 for 0 or 180 degrees and drops to 0.3 for 30 or 150 degrees [2], [3], [9].

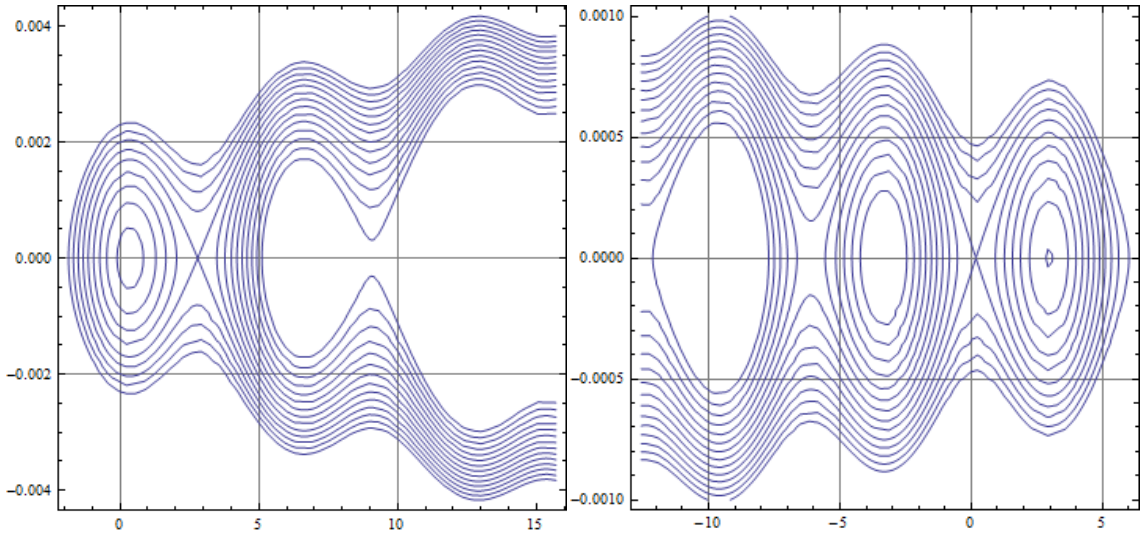

**Fig. 3:** Trajectories in **phase space** ($\phi, \tilde{p}/p$)($\phi, \tilde{p}/p$). x-axis in radian. The y-axis is the relative momentum deviation. The figure would be identical using the relative energy deviation. Accelerating bucket. Left: situation **below transition**, 20 degrees stable phase. The trajectories are travelled in the **anti-clockwise** direction. Right: situation **above transition**, 170 degrees stable phase. Trajectories travelled in the **clockwise** direction.

– The particles will occupy an area inside the bucket. We call this area the **bunch longitudinal emittance**. The RF voltage must be dimensioned to allow for capture and acceleration without loss. The bucket area must always be significantly larger than the bunch emittance. The ratio is called the **filling factor**.

## 1.4 Synchrotron tune spread and its consequences

The synchrotron motion in a non-accelerating bucket ($\phi_s = 0$) is described exactly by the pendulum system shown on Fig. 4.

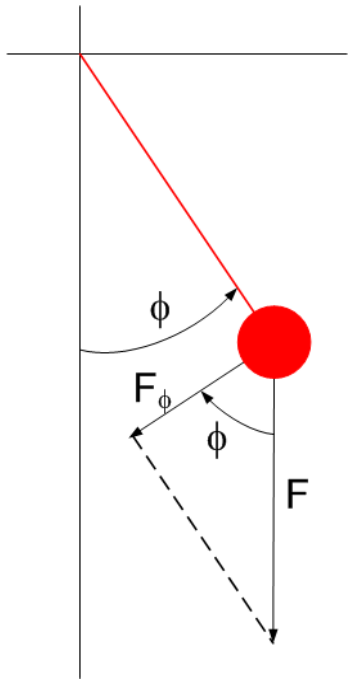

**Fig. 4:** Pendulum of mass $m$ and length $R$

We derive the equations of motion by writing the tangential part of Newton's equation

$$m\, j_\varphi = F_\varphi \quad , \tag{37}$$

$$-m\, R\, \frac{d^2\varphi}{dt^2} = m\, g\, \sin\varphi \quad , \tag{38}$$

$$\frac{d^2\varphi}{dt^2} + \frac{g}{R} \sin\varphi = 0 \quad . \tag{39}$$

Equation (39) is identical to Eq. (26) if $\phi_s = 0$ (non-accelerating bucket). In particular it contains the sine dependence that brings complexity to the synchrotron motion. It is well known that, for small amplitudes of oscillation, the phase space trajectory is a circle (if scaled correctly) and the period of oscillation is constant (after all, a pendulum has long been used to measure time). For larger amplitudes, however, the non-linearity has consequences: Fig. 5 shows increasing amplitudes and a bit of intuition will be required from the reader. When the amplitude gets large, the period increases, but the pendulum does not describe the phase space trajectory at constant speed anymore. It spends most of its time at the extremes of its oscillation. When the pendulum reaches maximal vertical position, it seems to hesitate before falling back on its downward swing. Try it… If the initial conditions place the pendulum at $\pi$ or $-\pi$, with zero speed, it will (at the limit) take an infinite time to make a complete oscillation. This is the equivalent of the separatrix.

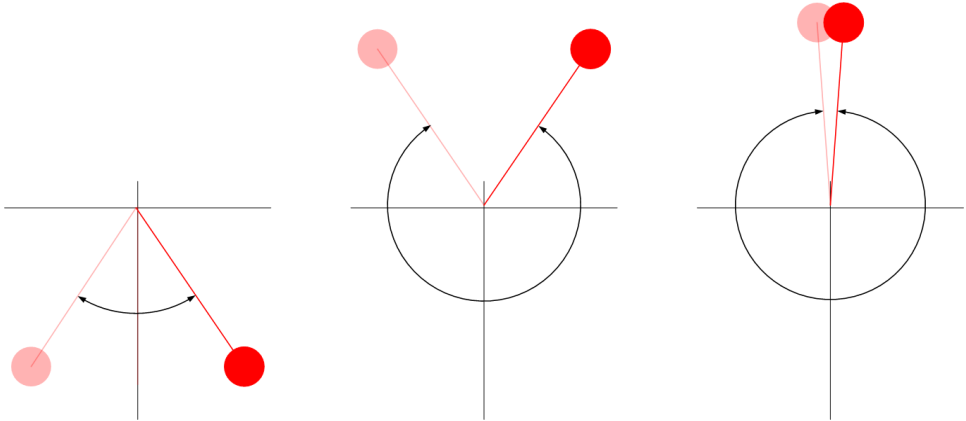

**Fig. 5:** Pendulum with increasing amplitudes of oscillation. When the extremes get close to $+-\pi$, the period becomes very large and the pendulum spends most of its time at the extremes of its oscillation.

From the pendulum we have learnt that

- The synchrotron frequency depends on the amplitude of the oscillation. Equation (30) applies to the centre of the bucket only and we will rename the zero-amplitude synchrotron frequency $\Omega_{s0}$.

- For larger amplitudes, the synchrotron frequency is smaller and finally drops to zero on the separatrix.

- Around the centre of the bucket (small amplitudes of oscillation) the particle travels at constant speed on the phase space trajectory (pure harmonic oscillator).

- But for larger amplitudes, the particle spends more time at the extremes of its oscillation (quasi-harmonic oscillator).

For $\phi_s = 0$ the frequency of synchrotron oscillation versus peak phase $\phi_{pk}$ between 0 and $\pi$, is [9]

$$\Omega_s(\varphi_{pk}) = \frac{\pi\, \Omega_{s0}}{2 \int_0^{\pi/2} \frac{du}{\left(1 - \sin^2\left(\frac{\varphi_{pk}}{2}\right) \sin^2 u\right)^{1/2}}} \quad . \tag{40}$$

It is plotted in Fig. 6, together with an approximation (very) valid for moderate amplitudes

$$\Omega_s(\varphi_{pk}) \approx \Omega_{s0}\left[1 - \left(\frac{\varphi_{pk}}{4}\right)^2\right] \quad . \tag{41}$$

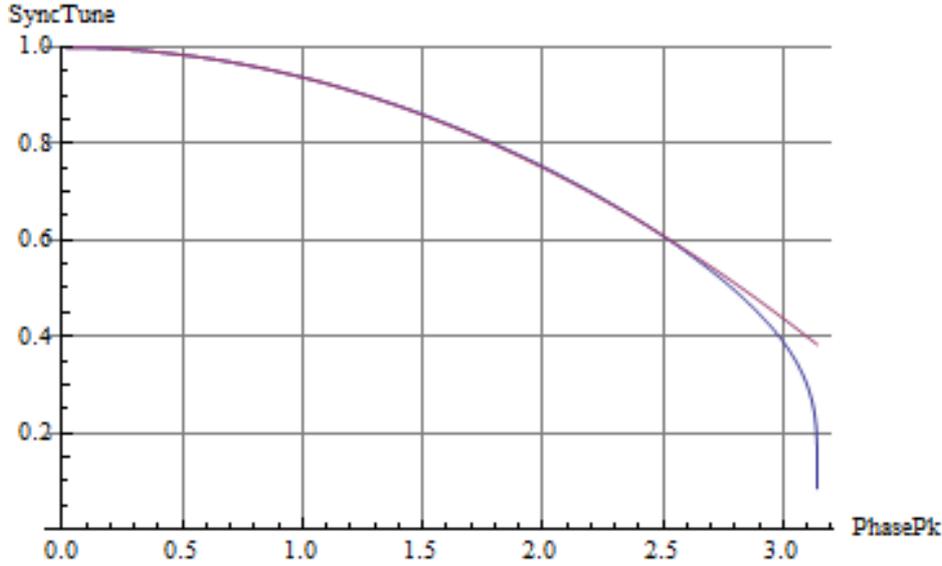

**Fig. 6:** $\Omega_s/\Omega_{s0}$ as a function of the maximum phase deviation in radian. Exact formula (bottom trace, blue) and approximation Eq. (41). Non-accelerating bucket ($\phi_s = 0$).

Analysis

- Given its length, the bunch will have a spread in the synchrotron tunes of the various particles. The longer the bunch, the larger the tune spread (for a given RF frequency).

- In hadron machines this tune spread will provide a stabilizing mechanism against coherent instabilities, called **Landau damping**.

- **Harmonic RF systems:** Adding an harmonic system (2 × or 4 × RF) we can shape the synchrotron tune vs. peak deviation curve. We may wish to increase the spread to increase Landau damping for stability (200/800 MHz systems in the SPS for example). Or we may wish to reduce the spread, to make the potential more linear and reduce the filamentation at injection (see below). Both are possible by adjusting the relative amplitude and phase of the fundamental and harmonic.

During filling, if the bunch is injected off-centred in the receiving bucket, its shape will be modified as the particles have different synchrotron frequencies: the trajectories in phase space will be travelled at different speed, fast for the particles around the centre of the bucket and slow for the ones injected close to the separatrix. Parts of the bunch will lag behind the core, resulting in **filamentation** in phase space (Fig. 7). After complete filamentation, the emittance will be much larger, filling the entire space within the blue trace in the simulation shown on Fig. 7.

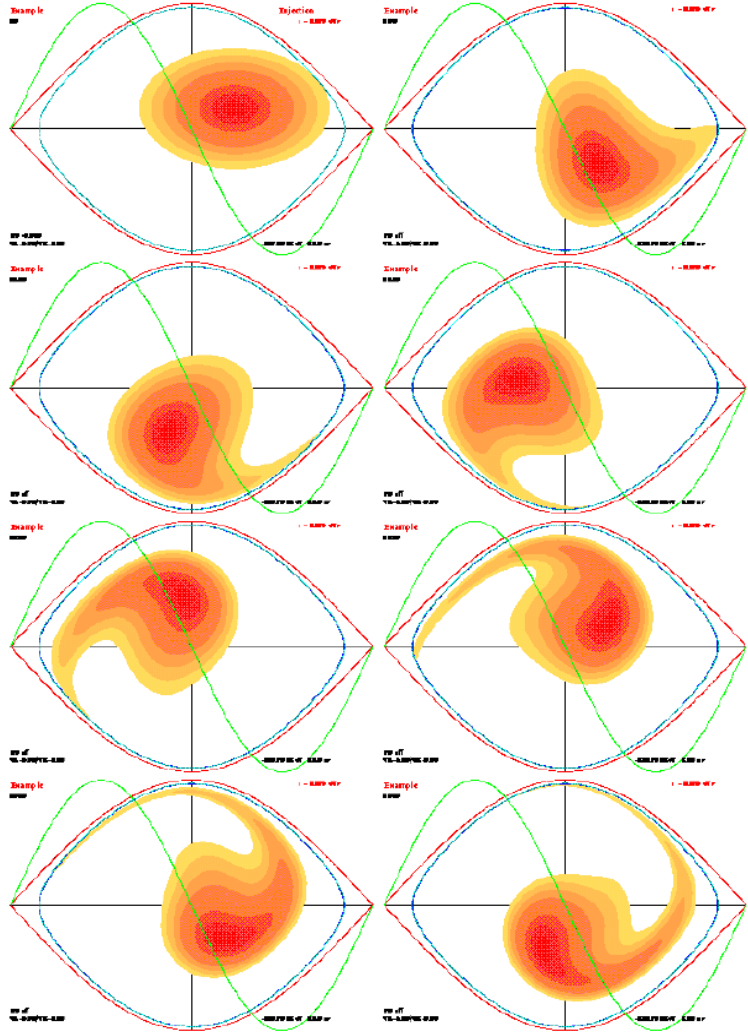

**Fig. 7:** Simulation of the filamentation at injection in the LHC bucket (phase space in [momentum, phase] units, above transition and thus clockwise displacement on the trajectories). The bunch is injected with a small phase/momentum error. The separatrix is in red. The evolution is left to right and top to bottom. After filamentation the bunch will fill the full area inside the blue contour resulting in an almost-full bucket. Courtesy of J. Tuckmantel.

## 1.5 RF capture optimization

We consider bunch-into-bucket transfer: the bunches must be transferred from the buckets of an injecting machine into the middle of the buckets in the receiving machine. In accelerator chains the optimal RF frequency tends to increase with energy so that the width of the receiving bucket is much smaller than the width of the injecting bucket if expressed in seconds. So the tolerance to phase errors is small. In the SPS–LHC case we transfer from a 200.4 MHz bucket into a 400.8 MHz bucket. We assume that the two RF systems are properly locked together. See Refs. [10] and [11] for technical details on RF synchronization between synchrotrons.

As the momentum and charge do not change in the transfer line, Eq. (5) requires

$$\rho_1 B_1 = \rho_2 B_2 \quad , \tag{42}$$

where index 1 and 2 refer to the two machines. Coarse matching of the two magnetic fields is first done with RF OFF: the beam is injected and the trajectories are measured on the first few turns in the receiving machine. Then using Eq. (16) one can derive the trim on $B_2$ that will centre the average beam trajectory.

Once the magnetic fields are matched we can switch the RF ON and move to the fine adjustments of the RF parameters of the receiving machine: frequency, phase, and voltage. The capture simulated on Fig. 7 is a catastrophe: the injected bunch density is colour-coded with dark red for the dense core and light yellow for the edges. It is injected with a phase and momentum error. Recall that momentum and frequency are equivalent at constant $B$ field, Eq. (18). As explained in the figure caption, it will filament and finally fill most of the bucket. As a result the bunch emittance has been blown up by a factor four to five during transfer and we end up with a significant population very close to the separatrix, ready to be lost out of the bucket at the weakest perturbation. So we need to inject in the centre of the bucket and that calls for fine-adjustment of the RF frequency and phase. With a longitudinal pick-up we can get a bunch profile at each passage. The measured bunch phase is the phase of the RF component of this bunch profile signal. The red dots (1, 2, 3) on Fig. 8 show the trajectory followed by the centre of the bunch after injection with a phase error (dot 1: correct momentum but displaced horizontally with respect to the bucket centre). The measured bunch phase would be a cosine at the synchrotron frequency. The green dots (a, b, c) correspond to an injection with the correct phase but with a momentum (or energy or frequency) deviation. The measured bunch phase will be a sine wave. By observing the bunch phase transient at injection we can derive the proper phase and frequency trims.

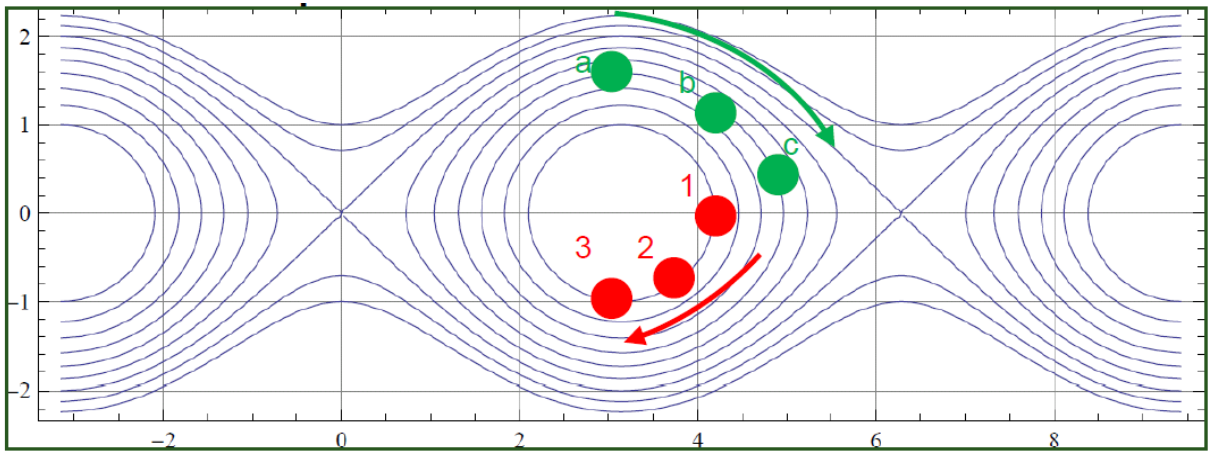

**Fig. 8:** Phase Space at injection. X-axis: Phase in radian. Y-axis: momentum, energy or frequency deviation from synchronism. The red dots (1, 2, 3) show the trajectory followed by the bunch centre for an injection phase error. The green dots (a, b, c) correspond to an injection momentum, energy, or frequency error.

Figure 9 shows an example pick-up signal. We observe the bunch profile during 1000 turns following injection. The phase of the centre (peak) of the bunch describes a sine wave, indicating an energy error. This bunch behaviour is called **dipole oscillation**: the bunch profile does not change but the phase of the centre of charge moves back and forth with respect to the stable phase. A frequency trim will cure the problem.

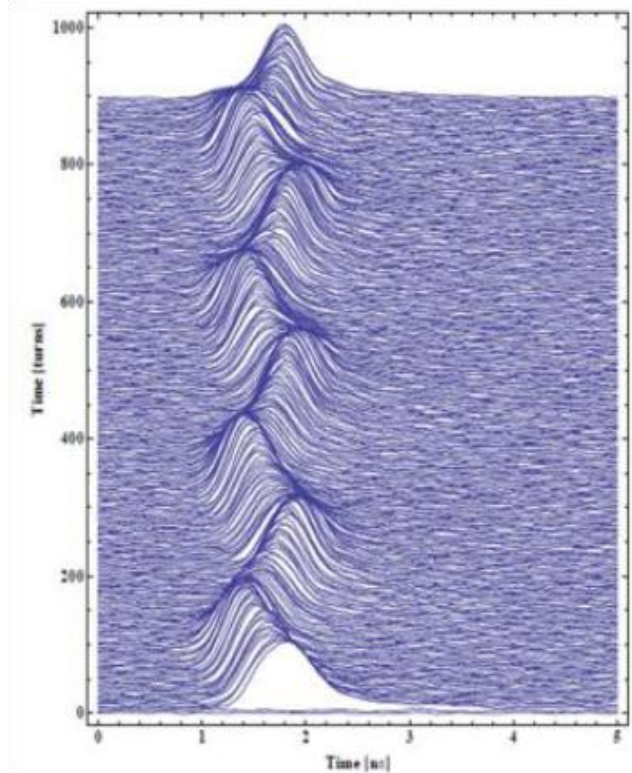

**Fig. 9:** Mountain range display: bunch profile measured turn after turn following injection. Horizontal axis in ns. Vertical axis in turn number (1000 turns total). The first few traces are recorded just before injection.

Be aware that, as the injector RF is locked to the receiving machine for transfer, a trim of the receiving RF frequency will also change the situation in the injector (small radial displacement at top energy and small change in the momentum at transfer). This may call for re-adjustment of the magnetic field in both machines. In theory, from a proper application of Eqs. (12) to (15) to the two machines, one could implement perfect energy matching in a single trial but I have observed that several iterations were always needed for a good result. The magnetic field and RF frequency are well adjusted when both first turn and beam circulating after injection transient, are centred.

Let us now consider matching the RF voltage at injection. Figure 10 shows the capture of a bunch (marked in red) with perfect phase and energy matching. The centre of the bunch falls in the middle of the bucket. The bunch has a non-zero length and therefore occupies an area defined by the phase space trajectories in the **injector**. But it is not matched to the phase space trajectories in the **receiving** machine (the voltage is too high). The particles of the bunch will follow these trajectories, resulting in the evolution shown on the figure: after one-quarter synchrotron period, the bunch length has been reduced (projection on the phase axis) and the momentum spread has been increased. We call this a **quadrupole oscillation**. It is a modulation of the bunch length (and momentum spread) at **twice the synchrotron frequency.** After filamentation the bunch emittance will be much increased and this must be avoided.

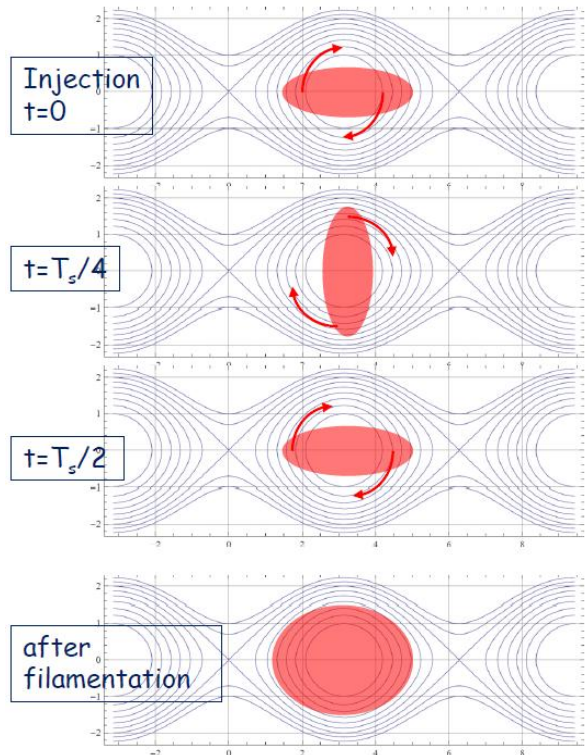

**Fig. 10:** Evolution in phase space. injection of a bunch (dark red) in the exact centre of the bucket but with phase space trajectories mismatched to the two-dimensional phase-momentum bunch profile. The result is a quadrupole oscillation at twice the synchrotron frequency and, after filamentation, significant emittance increase.

Figure 11 shows the time evolution of the bunch profile at injection with voltage mismatch. It corresponds to the phase space shown on Fig. 10. The voltage here is too high, resulting in a reduction of bunch length first. If the voltage was too low, the bunch length would first increase.

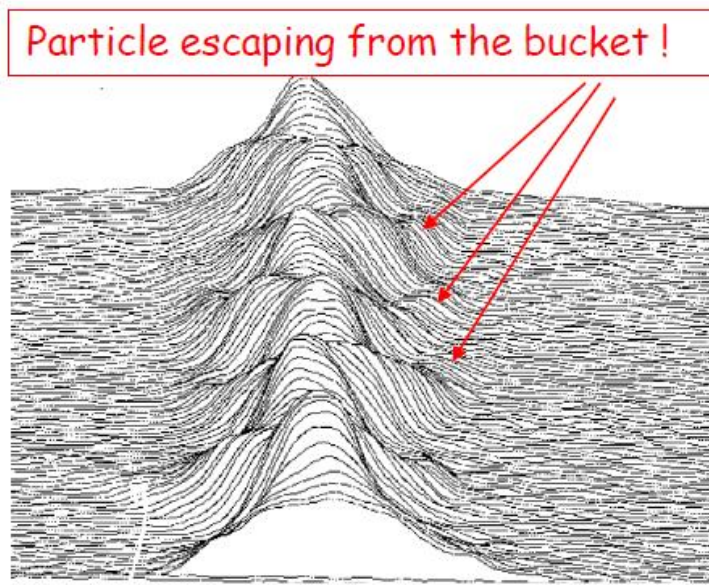

**Fig. 11:** Mountain range display: quadrupole oscillation at injection (plus some dipole and loss) indicating a voltage mismatch. Voltage too high.

Voltage matching is very easy. Feeding the longitudinal pick-up signal into a simple peak-detector, we get a monitoring of the bunch profile peak over time. In the absence of loss the product of bunch peak and length is constant. So quadrupole oscillations are clearly visible at the peak-detector output. Figure 12 shows this signal at the LHC injection for two different RF voltages, mismatched on the left and matched on the right.

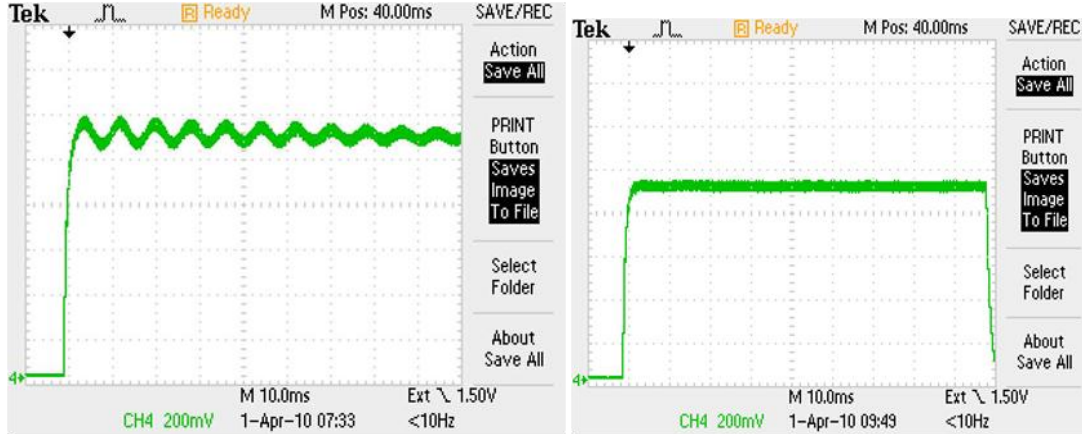

**Fig. 12:** Peak-detected pick-up signal showing quadrupole oscillation. Left: 8 MV. Right: 2.5 MV.

The RF voltage should be matched at capture to preserve the longitudinal emittance. In high-intensity machines, however, a larger voltage helps fight the effect of beam loading. If the emittance budget allows for some blow-up during transfer, we would then best capture with an higher-than-matched voltage. In 2010 we operated the SPS–LHC transfer with 3.5 MV while Fig. 12 shows that 2.5 MV would be matched best.

## 1.6 Radiation damping

When a relativistic charged particle is accelerated (meaning that its speed vector changes) it radiates energy at a rate proportional to the square of the accelerating force. In a circular accelerator the main accelerating force is the bending of the trajectory. For a particle moving at constant speed on a circular orbit of radius $\rho$, the power radiated is

$$P_\gamma = \frac{2}{3} r_0 E_0 c \frac{\beta^4 \gamma^4}{\rho^2} \quad , \tag{43}$$

where $r_0$ is the particle classical radius. The radiated energy must be compensated by the RF voltage. At 104.5 GeV/$c$ per beam, LEP required 3.66 GV RF. In the presence of significant radiation loss, the stable phase will not be zero (or 180 degrees) even if the magnetic field is constant. The energy lost by the synchronous particle must be compensated by a corresponding acceleration in the cavities. Notice that the radiated power increases sharply with energy (fourth power). During a synchrotron period in phase space, the mechanism will have a damping effect on the synchrotron oscillation of the non-synchronous particle: when its energy is larger than the synchronous energy it will radiate more and thereby lose part of the excess. Then in the bottom half of its synchrotron oscillation where its energy is lower than the synchronous energy, it will radiate less, thereby reducing its energy deviation (Fig. 13). This is modeled as a damping rate $\alpha_\gamma$ in the synchrotron motion equation

$$\frac{d^2 \tilde{\varphi}}{dt^2} + 2\alpha_\gamma \frac{d\tilde{\varphi}}{dt} + \frac{\Omega_{s0}^2}{\cos\varphi_s}\left(\sin\varphi - \sin\varphi_s\right) = 0 \quad . \tag{44}$$

The damping term is proportional to the derivative of $P_\gamma$ with respect to $\gamma$ and, after some manipulations, we get a simple and elegant approximated expression for the damping rate [2], [4], [5]

$$\alpha_\gamma \approx \frac{P_{\gamma s}}{E_s} \quad , \tag{45}$$

where $P_{\gamma s}$ is the power radiated by the synchronous particle. The damping time is thus the time that it would take for the synchronous particle to radiate out all its energy. Radiation damping is significant for circular electron accelerators and storage rings only because the radiated power scales as $\gamma^4$ and hadrons are not relativistic enough yet. In the LHC the radiation damping time is ~24 hours at 7 TeV/*c* and ~ 384 hours (more than two weeks) at the reduced 3.5 TeV/*c* used in 2010. Not much damping! The 7 TeV/*c* protons in the LHC have $\gamma$ ~ 7000 while the 100 GeV/*c* electrons and positrons in LEP had $\gamma$ ~ 200 000.

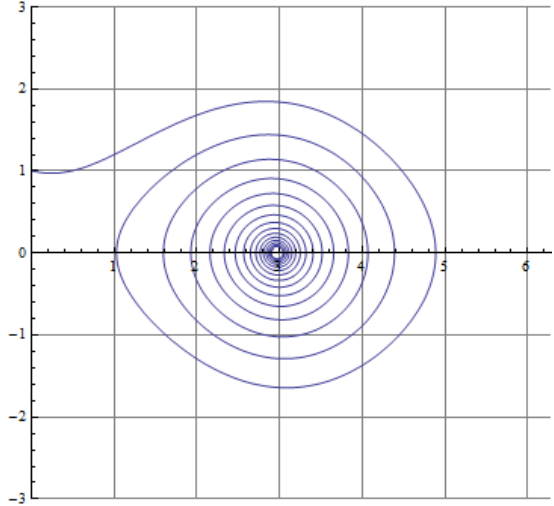

Fig. 13: Phase space trajectory of a non-synchronous particle with radiation damping. Evolution predicted from Eq. (44).

Where significant, radiation damping has a decisive impact on the bunch profile. With the damping introduced above, all non-synchronous particles would slowly spiral in phase space as shown on Fig. 13, converging to the centre of the bucket, resulting in zero longitudinal emittance and a point-like bunch. This is not the case. Radiation damping does indeed lead to very short, but not point-like, bunches in high-energy lepton storage rings. Electromagnetic radiation is emitted in quanta of discrete energy. In the phase-space representation, when a quantum is emitted, the momentum of the particle changes and it jumps on another, lower energy trajectory. This brings the bunch closer to the centre of the bucket if the quantum was emitted in the excess-energy part of the trajectory, but away from synchronism if emitted in the lower-energy part. This is similar to the classic statistical random walk process: at each trial we can take one step forward or backward. After a large number of trials, the average position is still zero but the variance keeps growing. The effect of many small jumps in phase space creates **diffusion**. The bunch length will be an **equilibrium** between the damping and the excitation due to the stochastic nature of the process. Refer to Refs. [4] and [5] for a detailed presentation. The bunch profile is also shaped by the radiation emission: for a given particle, emissions of successive quanta are independent. The central limit theorem states that, if a random variable is the sum of a large number of independent variables, its distribution becomes Gaussian no matter what the distribution of the individual random variables is. In high-energy lepton storage rings the bunch profile is indeed Gaussian and the distribution can be characterized by a single number (usually the variance $\sigma$ of the longitudinal bunch profile in either length or time, as measured by a longitudinal pick-up). That is not the case in hadron machines where the bunch profile depends greatly on the manipulations suffered in the acceleration chain.

## 1.7 Adiabatic evolution

So far we have considered the synchronous particle parameters ($p_s$, $E_s$, $\phi_s$) and the RF voltage as constant and have moved them out of the time derivatives. In an accelerator these parameters vary during the ramp. The voltage is matched to the injector at capture, then increased during the ramp to keep a sufficient bucket area with a non-zero (or 180 degrees) stable phase. The evolution is adiabatic if the relative variation of the synchrotron frequency in one synchrotron period is small

$$\left| \frac{\frac{d\Omega_{s0}}{dt}}{\Omega_{s0}} \right| \frac{2\pi}{\Omega_{s0}} \leq e^{-1} \quad . \tag{46}$$

In that case the particle stays on the same trajectory in phase space as this trajectory slowly adapts to the changing bucket. The Boltzman–Ehrenfest adiabatic theorem can be applied [2]: *"If (p,q) are canonically conjugate variables of an oscillatory system with slowly changing parameters, then the action integral, evaluated over one period of oscillation, is constant."*

$$I = \oint p\,dq = C \quad . \tag{47}$$

Applying this theorem to a closed trajectory in the longitudinal phase space we get the following relations describing the evolution of the maximum time $\Delta t_{pk}$ and energy $\Delta E_{pk}$ deviations in $E$–$t$ phase space, valid for adiabatic ramping (changes of $E_s$) and adiabatic voltage ($V$) variations:

$$\Delta t_{pk} \propto \frac{1}{f_{RF}} \sqrt[4]{\frac{|\eta|}{\gamma V \cos \varphi_s}} \quad , \tag{48}$$

$$\Delta E_{pk} \propto f_{RF} \sqrt[4]{\frac{\gamma V \cos \varphi_s}{|\eta|}} \quad . \tag{49}$$

Apart from the singularity at transition ($\eta = 0$), bunch length shrinks and energy spread increases with voltage increase and slow ramping (constant stable phase), if adiabatic. The effect is, however, moderate (fourth root). For the handling of the singularity at transition, see Ref. [2]. Figure 14 shows the evolution of the LHC bunch length (4 σ) during adiabatic manipulations: voltage increase before the start of the magnetic ramp (resulting in a sharp bunch shortening), followed by ramping.

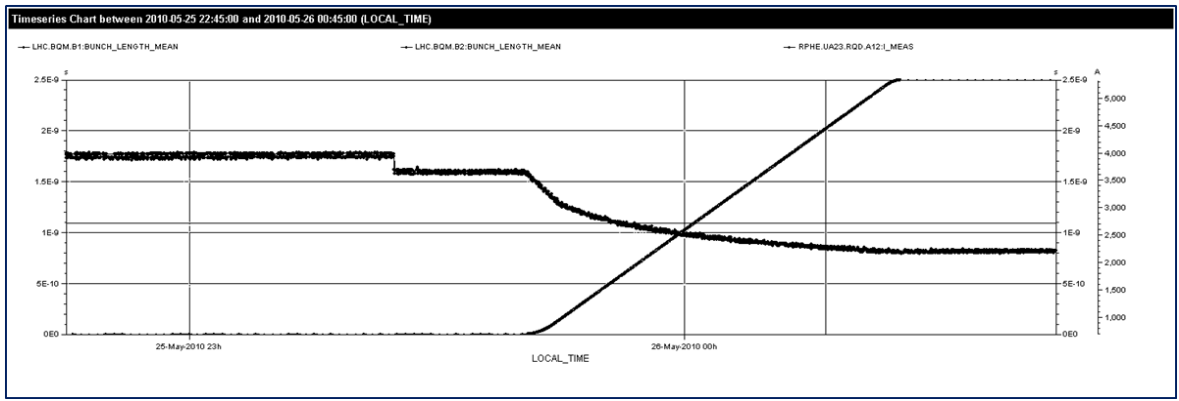

**Fig. 14:** Bunch length evolution for both beams in one early LHC ramp: capture at 450 GeV/*c* with 3.5 MV, voltage increase to 5 MV before start ramp then rise to 8 MV in the first part of the ramp (up to 3.5 TeV/*c*). Bunch length (4σ) evolution. Beam 1: 1.82 ns -> 1.61 ns -> 0.83 ns / Beam 2: 1.75 ns -> 1.58 ns -> 0.77 ns

Applying Eqs. (48) and (49) to the outer trajectory of the bunch in *E*–*t* phase space we conclude that **the longitudinal emittance (in eVs) remains constant during adiabatic evolution**.

## 2   Beam-based loops for synchrotrons

In this section, we study control loops that use signals from beam pick-ups, either longitudinal (beam phase) or transverse (beam position) and that act on all bunches.

### 2.1   Beam-phase loop

In the previous section we saw that phase/energy/voltage mismatch at capture will trigger dipole or quadrupole oscillations resulting in emittance blow-up after filamentation. In static conditions, the RF noise will excite the synchrotron oscillation of each particle individually, with the same result. In electron machines the synchrotron radiation provides a natural damping mechanism and will be sufficient in most cases except for the injection transient. In proton and ion machines there is no such natural damping. The bunch lengthening caused by the RF noise may lead to beam loss when particles reach the separatrix (major concern in colliders where beams are kept colliding for several hours). The beam-phase loop is designed to damp the dipole oscillation of the bunch.

Let us first consider the synchrotron oscillation in presence of a small $\delta\omega_{RF}$ modulation of the RF frequency. The kinematic relation for the RF phase at cavity crossing time, Eq. (20), becomes

$$\frac{d\tilde{\varphi}}{dt} = -2\pi h\ \tilde{f}_{rev} + \delta\omega_{RF}\ . \tag{50}$$

The first term is the effect of the momentum error and the second term is the RF frequency modulation. Following the derivation of the previous section, we get a modified linearized synchrotron motion equation

$$\frac{d^2\tilde{\varphi}}{dt^2} + \Omega_{s0}^2\ \tilde{\varphi} = \frac{d\delta\omega_{RF}}{dt}\ . \tag{51}$$

The phase of the beam is defined as the phase of the Fourier component of the beam current at the RF frequency. If the buckets are not evenly filled around the machine, the beam current will have a strong amplitude modulation at the revolution frequency. Its spectrum thus shows side-bands at $f_{rf} \pm nf_{rev}$. These must be filtered out of the beam phase signal to avoid exciting higher order coupled bunch dipole oscillations, ($n > 0$) with the phase loop.

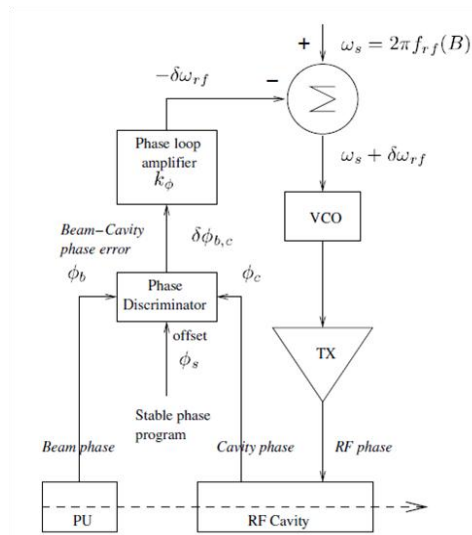

**Fig. 15:** Beam-phase loop

Consider the LLRF system shown on Fig. 15: a longitudinal pick-up provides a measurement of the beam phase that is compared to the phase of the cavity field (or to the vectorial sum of the cavity fields, with proper compensation for the time of flight, if the accelerator contains several cavities). The difference (beam–cavity phase error) minus the stable phase is used to correct the RF frequency via the phase loop amplifier. The inherent delays and bandwidth limitations in the beam-phase loop make it impossible to act bunch per bunch. We assume that the beam phase is, at each turn, averaged over all bunches in the machine. The simplest regulation is proportional only. $\omega_s$ is the RF frequency (in rad/s) derived from a measurement or an estimation of the magnetic field, Eq. (8). $\delta\omega_{rf}$ is the correction applied by the phase loop. We have

$$\delta\omega_{RF} = -k_\varphi \langle \tilde{\phi} \rangle \tag{52}$$

where the brackets stand for *averaging over all bunches*. The equation of synchrotron motion becomes

$$\frac{d^2 \langle \tilde{\varphi} \rangle}{dt^2} + k_\phi \frac{d \langle \tilde{\varphi} \rangle}{dt} + \Omega_{s0}^2 \langle \tilde{\varphi} \rangle = 0 \quad . \tag{53}$$

The synchrotron frequency is not changed but we have introduced the desired damping term. The above equation much resembles the damped equation resulting from radiation, Eq. (44). However, while radiation damps each particle individually, the LLRF phase loop can only act on the average dipole oscillation of all bunches. It must be fast compared to the filamentation time in order to damp phase and energy errors at injection before significant emittance blow-up. Figure 16 illustrates the action of the phase loop in normalized phase space. It considers injection in a non-accelerating bucket above transition. The point-like test bunch is injected at point (0, 1) in the normalized phase space, corresponding to a -π phase error and a momentum error equal to one half the bucket half-height. (The bucket is shown at the top left.) Displayed are the evolutions without phase loop (top right) resulting in no capture, and with phase loop for two different loop gain settings. These plots are somewhat confusing however: recall that **the phase loop does not displace the beam**. It changes the RF phase and frequency to jump the bucket onto the injected bunch. In Fig. 16 the phase loop actually **displaces the axis to bring the (π, 0) point right on the beam**. This explains why a phase loop can be much faster than the synchrotron period while remaining adiabatic.

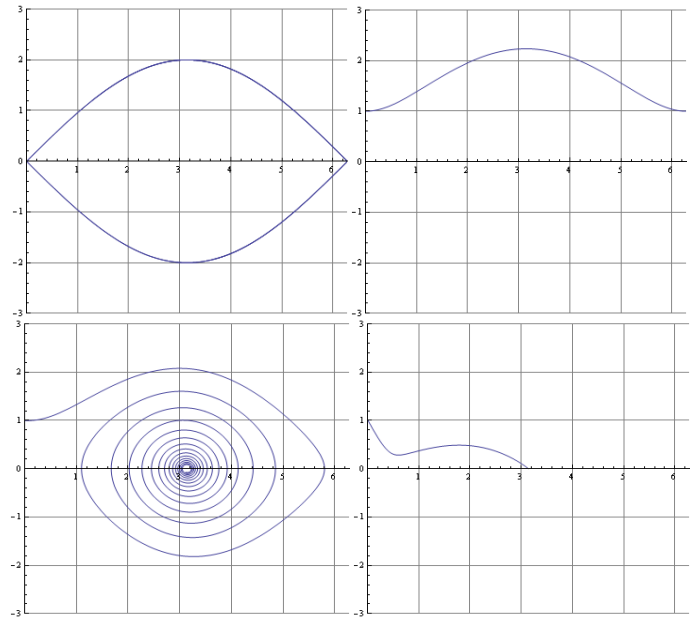

**Fig. 16:** Injection transients in **normalized phase space** ($\phi$, $1/\Omega_s \, d\phi/dt$) above transition for synchronous phase 180 degrees. RF bucket (top left). Injection of a point-like bunch with phase and energy error point (0, 1). The evolution without phase loop is shown on the top right: the bunch surfs over the bucket and is not captured. The bottom two traces are with phase loop on at low gain (left) and high gain (right): the bunch is captured.

Figure 17 shows the fast damping of the LHC injection error, achieved with the phase loop.

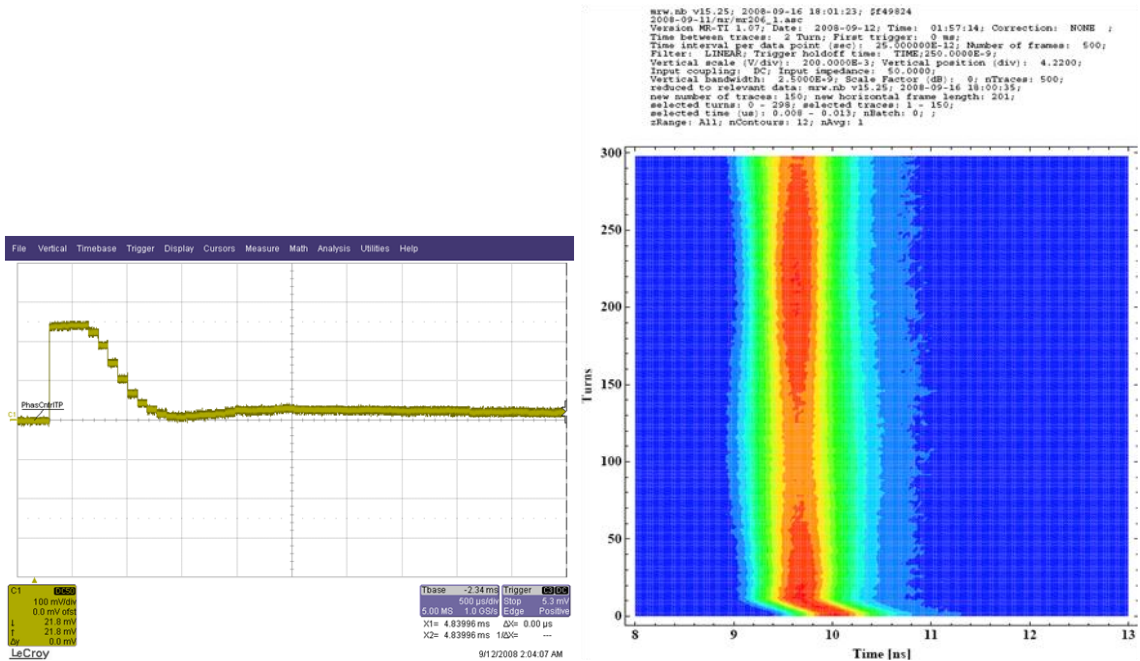

**Fig. 17:** Damping of the phase error at injection into the LHC. Single bunch (89 μs revolution period). Left: beam–cavity phase error, 500 μs/Div. After a four-turn latency, the error is brought to zero in about ten turns. Right: mountain range of the bunch at injection showing the fast damping of the phase error. The bunch profile is colour-coded. Notice the quadrupole oscillation caused by a voltage mismatch.

After injection, in quiet conditions, the phase loop is very efficient in fighting the effect of RF noise. This is important in hadron colliders where beams must be kept for several hours with minimal emittance blow-up and there is no natural damping. RF phase noise is more damaging than amplitude noise as the synchrotron phase is practically zero or 180 degrees in hadron colliders. The effect of RF noise can be observed by monitoring the bunch length. Figure 18 shows the clear correlation between bunch lengthening and phase loop gain in the LHC. A strong phase loop was also essential in the SPS when used as p–pbar collider. Be aware that the phase loop is useful only if the bunch fills a small portion of the bucket. With a full bucket any injection error or RF noise will result in particles escaping.

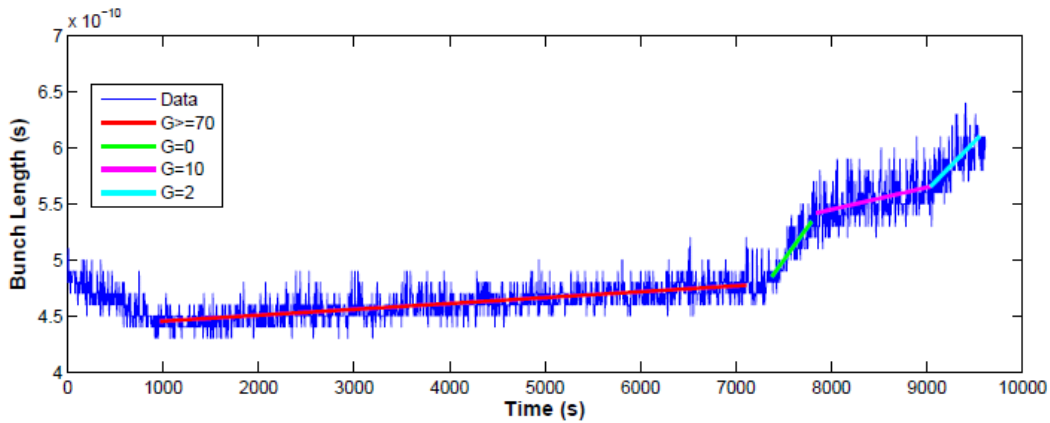

**Fig. 18:** LHC single bunch at 3.5 TeV/*c*, 1E10 p. Evolution of bunch length with time while varying the phase loop gain [12].

How does the phase loop act on bunch lengthening? The explanation is that the phase loop reduces the phase noise in the cavity sum signal in the synchrotron band. In this band, if the bunch length is short compared to the bucket width, the beam gives a coherent response that is measured by the loop and damped via the modulation of the VCO. Outside the synchrotron band the phase loop does nothing — except inject noise in the cavity, as there is no response from the beam. Note that it acts on the first synchrotron band only. Figure 19 confirms this analysis. It shows the phase noise Power Spectral Density (PSD) in one LHC cavity for varying phase loop gain. Notice the notch at the synchrotron frequency, whose depth depends on the loop gain.

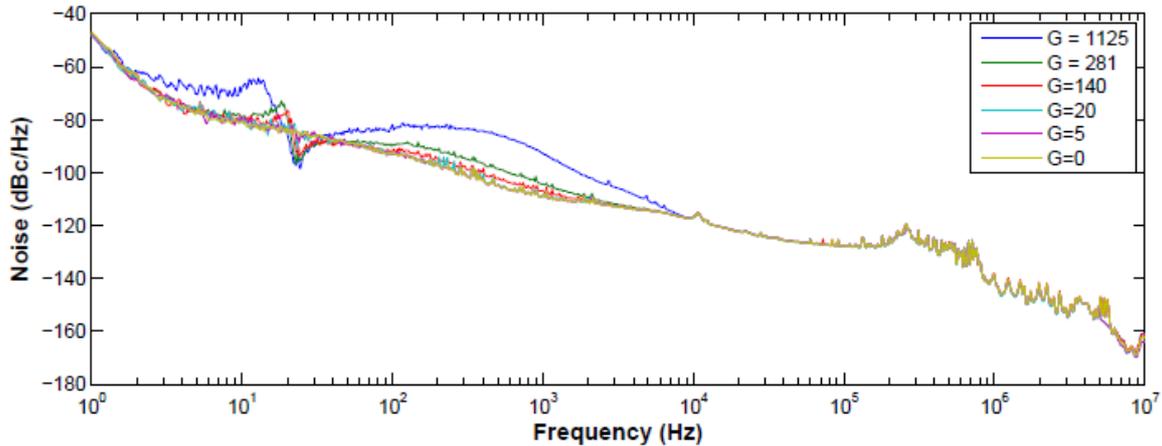

**Fig. 19:** Single-sideband phase noise PSD in dBc/Hz in an LHC cavity with circulating 3.5 TeV/*c* bunch, for various phase loop gains (in s$^{-1}$). The synchrotron frequency is ~ 24 Hz [12]

What is the optimal gain of the phase loop? The previous analysis suggests no limit. In a real implementation the closed loop will have a delay and that will limit the gain. A comprehensive treatment of the low-level loops in the presence of long delays can be found in Ref. [13].

We now need to introduce a bit of discipline: for the phase loop, the beam is the master. The RF will do its best to please it. If there is an energy error at injection, the RF will change its frequency. If there is noise in the cavity in the delicate synchrotron band the RF will be modulated to minimize this noise. That will preserve longitudinal emittance but — it is not a stand-alone solution because:

– If there are several injections, the RF must be restored to an injection frequency after transient to prepare for the next injection.
– When we start ramping, the RF must track the magnetic field to keep the beam centred.
– If we transfer to another machine, the RF must be synchronized to the buckets of the receiving machine.
– There is no mechanism to keep the beam centred.

*Solution:* We will add a slower loop that will **discipline** the beam. It will be gentle enough so that it does not perturb the all-important phase loop. In the context, gentle means **adiabatic** as defined in the previous section. There are several options and two big classics:

– The **radial loop**: we slowly adjust the RF to keep the beam centred as measured in one or several transverse pick-ups.
– The **synchro loop** with frequency program reference: we keep the RF softly locked onto a synthesizer whose frequency tracks the magnetic field to keep the beam centred.

We will now present these two options.

## 2.2 Radial loop

Figure 20 shows the combination of phase and radial loop. One (or several) transverse pick-ups provide a measurement of the beam radial position error that is used to correct the frequency via the radial loop amplifier.

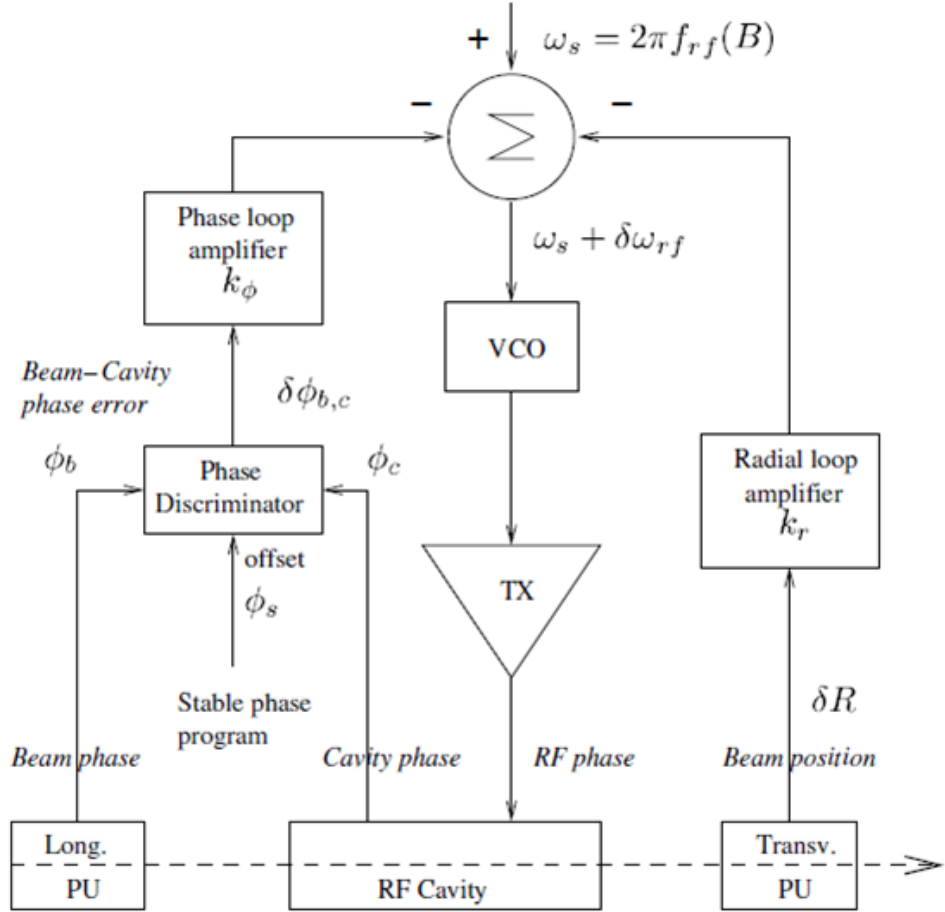

Fig. 20: A classic combination for proton and ion synchrotrons: phase loop and radial loop

The simplest regulation is proportional only. We have

$$\delta\omega_{RF} = -k_\phi \langle \tilde{\varphi} \rangle - k_R \langle \tilde{R} \rangle \quad . \tag{54}$$

At constant magnetic field, the radial displacement is proportional to the momentum deviation. From Eq. (12)

$$\frac{\langle \tilde{R} \rangle}{R} = \frac{1}{\gamma_t^2} \frac{\langle \tilde{p} \rangle}{p} \tag{55}$$

and the frequency trim becomes

$$\delta\omega_{RF} = -k_\phi \langle \tilde{\varphi} \rangle - k_R \frac{R}{\gamma_t^2 p_s} \langle \tilde{p} \rangle \quad . \tag{56}$$

Using Eqs. (51) and (23), we get, after linearization

$$\frac{d^2\langle\tilde{\varphi}\rangle}{dt^2} + k_\phi \frac{d\langle\tilde{\varphi}\rangle}{dt} + \left(\Omega_{s0}^2 + k_R \frac{qV\cos\varphi_s}{2\pi p_s \gamma_t^2}\right)\langle\tilde{\varphi}\rangle = 0 \quad . \tag{57}$$

Analysis:

- The radial loop does not provide damping. It only increases the frequency of oscillation. The damping is provided by the phase loop.

- The phase loop/radial loop tandem has very good behaviour during transients. At injection, for example, if the beam is injected with a phase and energy error the first turn will be on an off-centred orbit and the beam will see a non-zero RF voltage. The phase loop reacts in a few turns and the RF jumps on the bunch, thereby preventing emittance blow-up. Thereafter the radial loop will slowly modify the beam energy to drive it back to the centre orbit.

- In the system shown on Fig. 20, the stable phase $\phi_s$ must be subtracted from the beam–cavity phase error measured by the phase discriminator. It is computed from a measurement of the RF voltage and the momentum, Eq. (11). An error in the stable phase computation will introduce a radial displacement of the beam if the phase-loop amplifier is DC-coupled. To avoid this, the phase loop can be AC-coupled. This method is used in the PS accelerator at CERN [14].

- If one neglects the delays, the combination of phase loop/radial loop is unconditionally stable. A comprehensive treatment of the low-level loops in the presence of delays can be found in Ref. [13]. Other interesting references are [15], [16] (application to the CERN PS during the 1970s), and [17] (application to the CERN PS Booster).

- The sign of the radial gain must be changed at transition because $\cos\phi_s$ changes sign. It must be positive below transition and negative above.

- The radial loop is very efficient for reducing the effect of frequency errors on the radial position. Let us assume a small error $\delta\omega$ in the RF frequency. From Eq. (17) we derive the resulting error on the radial position (without radial loop)

$$\frac{\delta R}{R} = \left(\frac{\gamma^2}{\gamma_t^2 - \gamma^2}\right)\frac{\delta\omega}{\omega} \quad . \tag{58}$$

At transition, the sensitivity becomes infinite! With the radial loop the effect is limited by the closed loop gain, that cannot be too large in order to remain adiabatic. **The radial loop is required to cross transition.**

- The radial loop couples the transverse and longitudinal planes using transverse measurements to estimate momentum and correct the frequency. This causes problems: transverse betatron oscillations are interpreted as momentum error. This effect can be minimized by using two pick-ups at 180 degrees in betatron phase.

- The radial loop typically looks at one or few PUs only. It centres the beam in one location only, instead of centring the average orbit.

## 2.3 Synchro loop

Instead of monitoring the beam radial position we can lock the RF frequency on an external reference via a synchro loop. This loop must have a much smaller gain than the phase loop to guarantee adiabaticity. Figure 21 shows the combination of phase and synchro loop. We measure the phase of the RF (or beam) and compare it with the reference generator. This error is used to correct the RF frequency via the synchro loop amplifier. The reference generator will be set at the injection frequency during filling, then follow the frequency program during ramping, Eq. (8), and finally be locked to the receiving machine if needed for transfer.

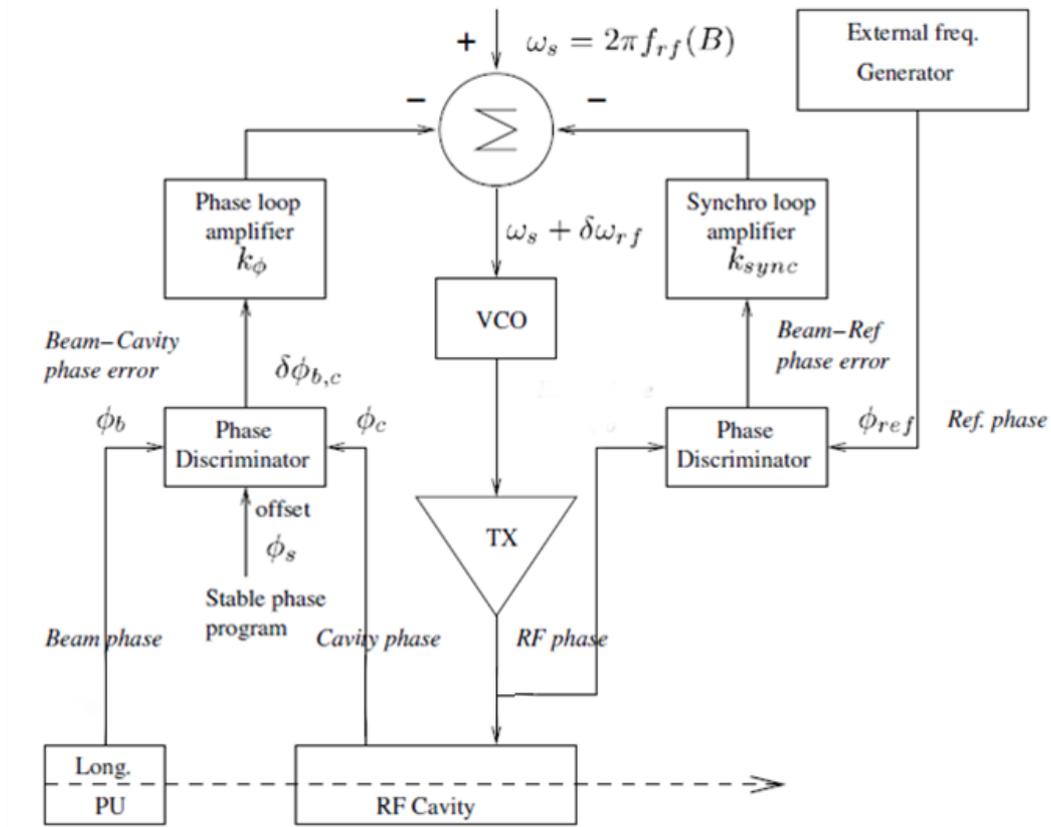

**Fig. 21:** Another classic combination for proton and ion synchrotrons: phase loop and synchro loop

The overall system is described by a third-order differential equation. Its analysis is easier if we use Laplace transforms. First the beam transfer function (undamped resonator at $\Omega_{s0}$), from Eq. (51)

$$\tilde{\varphi}(s) = \frac{s}{s^2 + \Omega_{s0}^2} \delta\omega_{RF}(s) = B_\varphi(s) \quad . \tag{59}$$

The open-loop transfer function from $\delta\omega_{RF}$ to $\delta\phi_{RF}$ (the phase modulation applied to the RF) with phase loop closed but synchro loop open is

$$H_{ol}(s) = \frac{\delta\phi_{RF}(s)}{\delta\omega_{RF}(s)} = \frac{1}{1 + k_\varphi B_\varphi(s)} \frac{1}{s} = \frac{s^2 + \Omega_{s0}^2}{s\left(s^2 + k_\varphi s + \Omega_{s0}^2\right)} \quad . \tag{60}$$

If the synchro loop amplifier is proportional only, the response gives too low a phase margin. It can be corrected by using a phase advance network as corrector in the synchro loop amplifier

$$H_{sync}(s) = k_{sync} \frac{1 + a\tau s}{1 + \tau s} \quad . \tag{61}$$

Figure 22 shows the Nyquist plot without and with the corrector. The parameters $a$ and $\tau$ can be optimized using classical controls theory to provide the usual 60 degrees phase margin. They will depend on the synchrotron frequency and, if this one varies much during the acceleration, the corrector parameters must track.

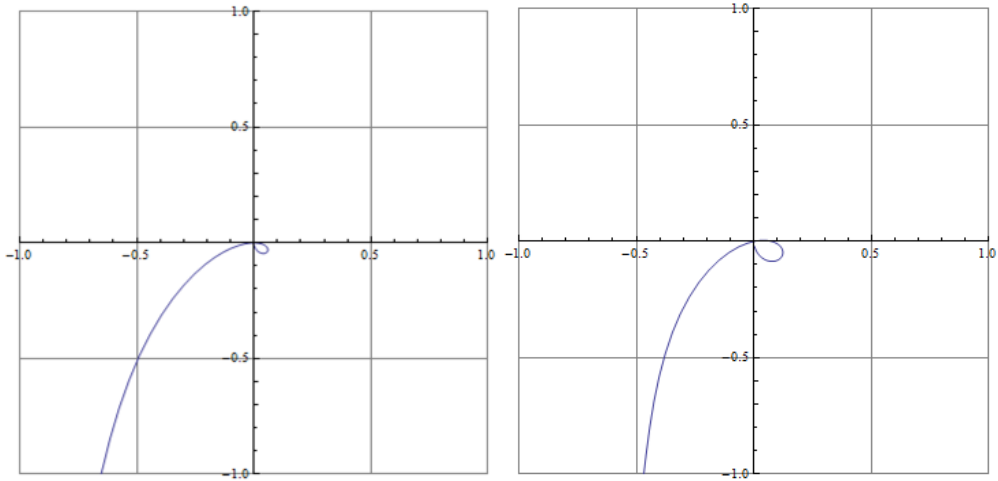

**Fig. 22:** Nyquist plots of $H_{ol}(j\omega).H_{sync}(j\omega)$ without phase advance network (left) and with correction optimized. Phase margin increased from ~45 to >60 degrees.

Analysis:

- The synchro loop must respond in a time that is larger than the synchrotron period to remain adiabatic and to avoid exciting the beam with noise around the synchrotron frequency.

- The loop can be used throughout the acceleration cycle: it is set at the injection frequency during filling, then ramped following either a measurement of the magnetic field (SPS proton for LHC) or a function if the magnetic field is well modelled (LHC).

- If the accelerator is an injector, the loop can remain in use while rephasing takes place locking the reference generator on the receiving machine RF (SPS proton for LHC).

- If it is a collider, using a single reference for both rings from filling to physics, makes the beam cross in the correct position from injection on (LHC).

- If needed, a slow measurement of the orbit displacement using all ring transverse horizontal PUs can feed back on the reference frequency (LHC real-time orbit correction).

- Limitation: It is **not practical if transition is crossed during the acceleration**. Equation (58) shows that, near transition, a very small frequency error will cause a large radial displacement. **The radial loop is preferred if the acceleration cycle crosses transition.**

- Limitation: **The range of the phase discriminator is 360 degrees only**. And the synchro loop is much slower than the phase loop to remain adiabatic. At injection, with large offsets (energy error or stable phase offset), the phase discriminator may reach its limits before the loop is locked. It will later lock with an error of one or several RF periods. This is not acceptable if the filling requires successive injections. In the LHC the drifts are very small and the offsets do not result in a synchro transient exceeding ± 180 degrees. The situation is different in the SPS where analog electronics is still in use and offsets vary much more from injection to injection. The problem has been solved by implementing the synchro loop at a sub-harmonic of the RF (40 MHz = fifth sub-harmonic). The 360 degrees range at the sub-harmonic virtually implements a multi-period phase range at the RF frequency.

The phase loop/synchro loop tandem performs very well in the LHC. We inject at 450 GeV/$c$, well above transition (~ 50 GeV/$c$). Figure 23 shows the injection transients: phase loop error on the left and synchro loop error on the right. The time is scaled in turns. Notice the different scales. The fast phase loop 'jumps' the RF on the beam in about ten turns, nulling any phase and energy error. In the presence of an energy error, the RF frequency will thus be driven away from the reference frequency as it follows the off-energy beam. The result is the observed linear phase ramp in the synchro loop error. After about three hundred turns (27 ms), the synchro loop has reacted slowly and brought the RF and beam back to the reference. Figure 23 is a screenshot of the display used by the LHC operation to monitor LLRF injection transients.

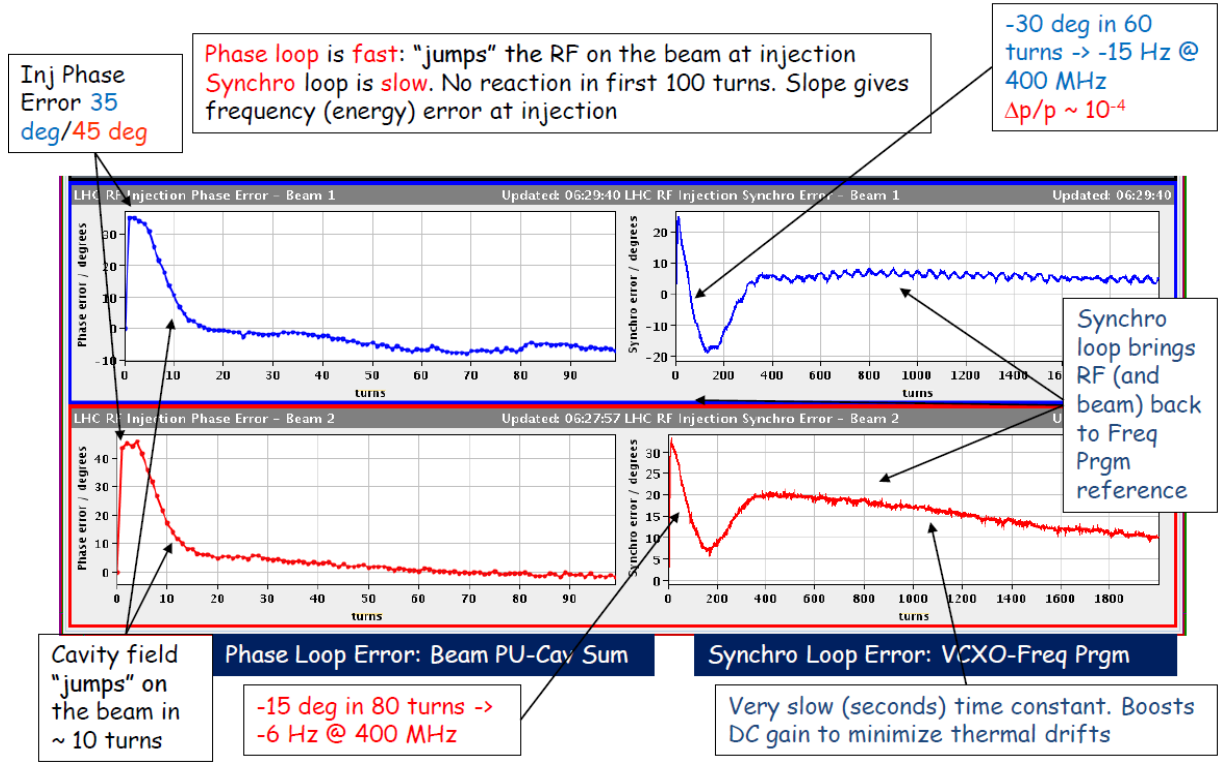

**Fig. 23:** LHC phase loop (left) and synchro loop (right) injection transients. Beam 1 (top) and beam 2 (bottom). Horizontal axis in turns (89 μs revolution period). Notice the different scales. Synchrotron frequency around 60 Hz.

Instead of using the RF phase one could implement a synchro loop comparing the beam phase with the reference generator. This adds a complication if the beam intensity covers a wide dynamic range: the design of a zero phase-shift limiter. The dynamics are similar but different. Refer to Refs. [14] and [17] for details.

**Appendix A: Relations between $E$ (total energy), $E_0$ (rest energy), $p$ (momentum), $v$ (speed), $\gamma$ and $\beta$**

$$\beta = \frac{v}{c}$$

$$\gamma = \frac{E}{E_0}$$

$$\beta = \sqrt{1 - \frac{1}{\gamma^2}}$$

$$\gamma = \frac{1}{\sqrt{1-\beta^2}}$$

$$E^2 = E_0^2 + c^2 p^2$$

$$\frac{dE}{dp} = \frac{pc^2}{E} = v$$

$$\beta = \frac{1}{\sqrt{1 + \left(\frac{E_0}{cp}\right)^2}}$$

$E_{0p} = 938.26$ MeV   proton rest energy

$E_{0e} = 0.511$ MeV   electron rest energy